# Direct visualization of non-equilibrium heat extraction by a metasurface absorber

Shu Watanabe,[1] Robert E. Simpson,[2] Takuo Tanaka,[3-6] and Wakana Kubo[1*†]

[1] *Department of Electrical Engineering and Computer Science, Tokyo University of Agriculture and Technology, 2-24-16 Naka-cho, Koganei-shi, Tokyo 184-8588, Japan*
[2] *School of Engineering, University of Birmingham, Edgbaston, Birmingham, B15 2TT, UK.*
[3] *Innovative Photon Manipulation Research Team, RIKEN Center for Advanced Photonics, Wako, Saitama 351-0198, Japan.*
[4] *Metamaterials Laboratory, RIKEN Cluster for Pioneering Research, Wako, Saitama 351-0198, Japan.*
[5] *Department of Physics, Faculty of Science, Gakushuin University, Tokyo 171-8588, Japan.*
[6] *Institute of Post-LED Photonics, Tokushima University, Tokushima 770-8560, Japan.*

Metasurface heat absorption is conventionally characterized by far-field infra-red spectroscopy under equilibrium conditions. The approach provides information on the surface emissivity but does not directly reveal heat transferred by electromagnetic absorption. To understand the resulting temperature change, however, the associated heat transfer must be considered. Here, we directly observe heat extracted by electromagnetic absorption using infrared thermography. By monitoring the cooling dynamics of a copper plate placed in contact with a metasurface-integrated thermoelectric device, we show that the copper plate cools faster than one placed on an unstructured planar surface (control). This result provides direct evidence of heat transfer from the adjacent solid to the metasurface; revealing a metasurface-mediated electromagnetic absorption heat transfer mechanism. These findings provide new insight into nonequilibrium thermal processes enabled by metasurface absorbers.

Metasurface absorbers provide a versatile platform for tailoring electromagnetic absorption over designed spectral ranges.[1-4] Despite extensive studies of their optical properties, their thermal response has been characterized mainly through far-field measurements of absorption spectra under equilibrium conditions.[5-10] While such measurements provide information on radiative emission, they do not directly reveal how absorbed electromagnetic energy is transferred to the surrounding medium.[11-14] Such measurements are commonly interpreted through Kirchhoff's law of thermal radiation, which relates absorptivity and emissivity under conditions of thermal equilibrium.

Under nonequilibrium conditions, however, electromagnetic absorption can be accompanied by heat transfer from the surrounding medium and, therefore, Kirchhoff's Law of Thermal Radiation alone is insufficient to describe the system. We have previously used this concept to demonstrate nonradiative cooling using metamaterial thermoelectric conversion, in which heat absorbed by a metasurface is transported outside through thermoelectric conversion, thus lowering the temperature inside a sealed space.[15-22]

In our previous experiments, temperature changes in the surrounding environment were monitored using thermistors placed near the electrodes of the thermoelectric device. However, these measurements provided only spatially averaged information, leaving heat transfer from the surrounding medium to the metasurface indirectly inferred. However, mapping the spatial temperature distribution would provide the direction and extent of the heat flow, which is important for understanding how to design systems with optimal thermal coupling and heat extraction efficiencies.

In this study, we investigate heat extraction associated with electromagnetic absorption using infrared thermography. By monitoring the temperature evolution of a metallic object in contact with the metasurface, we resolve heat transfer from the adjacent solid to the metasurface. This direct observation of heat transfer mediated by electromagnetic absorption is inaccessible by methods relying on equilibrium conditions, such as far-field IR spectroscopy. Indeed, this work provides insight into thermal energy transfer in out-of-thermal-equilibrium metasurface absorbers.

To visualize the heat extraction associated with electromagnetic absorption, we fabricated a metasurface absorber on a polished Cu substrate, as illustrated in Figs. 1(a)-1(c). The metasurface consisted of an Ag film, a calcium fluoride ($CaF_2$) spacer, and Ag disk. A total of 16 domains with dimensions of 1.5 mm × 1.5 mm were fabricated using electron-beam lithography and vacuum evaporation techniques. The Ag disk diameter, thickness, and periodicity were 3.5 µm, 100 nm, and 5 µm, respectively, while the thicknesses of the Ag film and $CaF_2$ spacer were 150 nm and 140 nm. Figure 1(d) shows the absorption spectrum of the metasurface. A pronounced absorption peak was observed at 10.5 µm, originating from the magnetic resonance mode of the structure.[23,24] Fig. 1(d) also shows the blackbody radiation spectrum calculated at 385 K, corresponding to the initial temperature of the Cu plate, which substantially overlaps the absorption band of the metasurface.

The metasurface was fabricated on the Cu electrode. For the remainder of this article, we refer to this as the MS-Electrode. The MS-Electrode was adhered to a thermoelectric device, as illustrated in Figs. 1(a) and 1(b). The thermoelectric device consisted of eight *p*-*n* thermoelectric pairs composed of *p*-type antimony telluride ($Sb_2Te_3$) and *n*-type bismuth telluride ($Bi_2Te_3$), with a Seebeck coefficient of $\pm$ 76 $\mu V \cdot K^{-1}$. A bare Cu electrode was attached to the opposite side of the device. For comparison, a device employing Cu electrodes without metasurfaces was also prepared as a control reference.

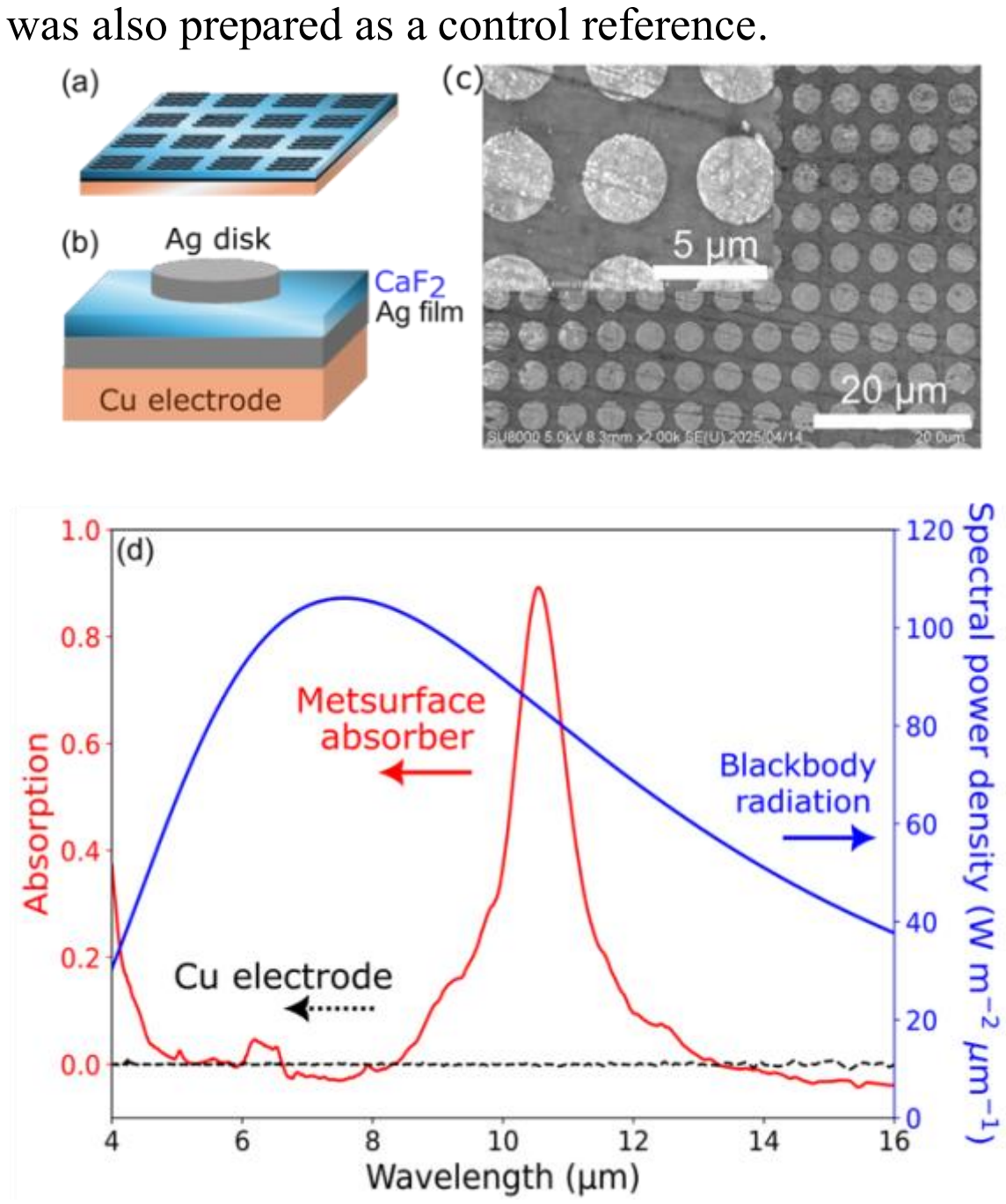

Figure 1. (a) Schematic of the MS-Electrode and (b) unit cell of the metasurface absorber. (c) Top-view SEM image of the fabricated metasurface absorber. The inset is a magnified view of the metasurface unit cell. Scale bars correspond to 20 and 5 µm, respectively. (d) Measured absorption spectra of the metasurface absorber (red solid line) and bare Cu electrode (black dashed line), together with the blackbody radiation spectrum at 383 K (blue line).

Figure 2(a) shows the experimental setup used to visualize the heat extraction process. The metasurface and control thermoelectric devices were placed on a thermally insulated stage, and Cu plates were brought into direct contact with the upper electrodes of the devices. To improve thermography measurements, a polyimide film with a thermal conductivity of 0.12 $W \cdot m^{-1} \cdot K^{-1}$, dimensions of 10 mm × 10 mm, and a thickness of 0.15 mm was attached to the top surface of each Cu plate to suppress reflection from the Cu surface.

The samples were heated using a Peltier heater set to 383 K for 30 s and subsequently allowed to cool after removal of the heater. The temporal evolution of the surface temperatures was monitored by infrared thermography (FLIR, FLIR One Pro), while the electrical output of the thermoelectric devices was measured simultaneously using multimeters (KEITHLEY, DMM-6500). The experiment was conducted in air.

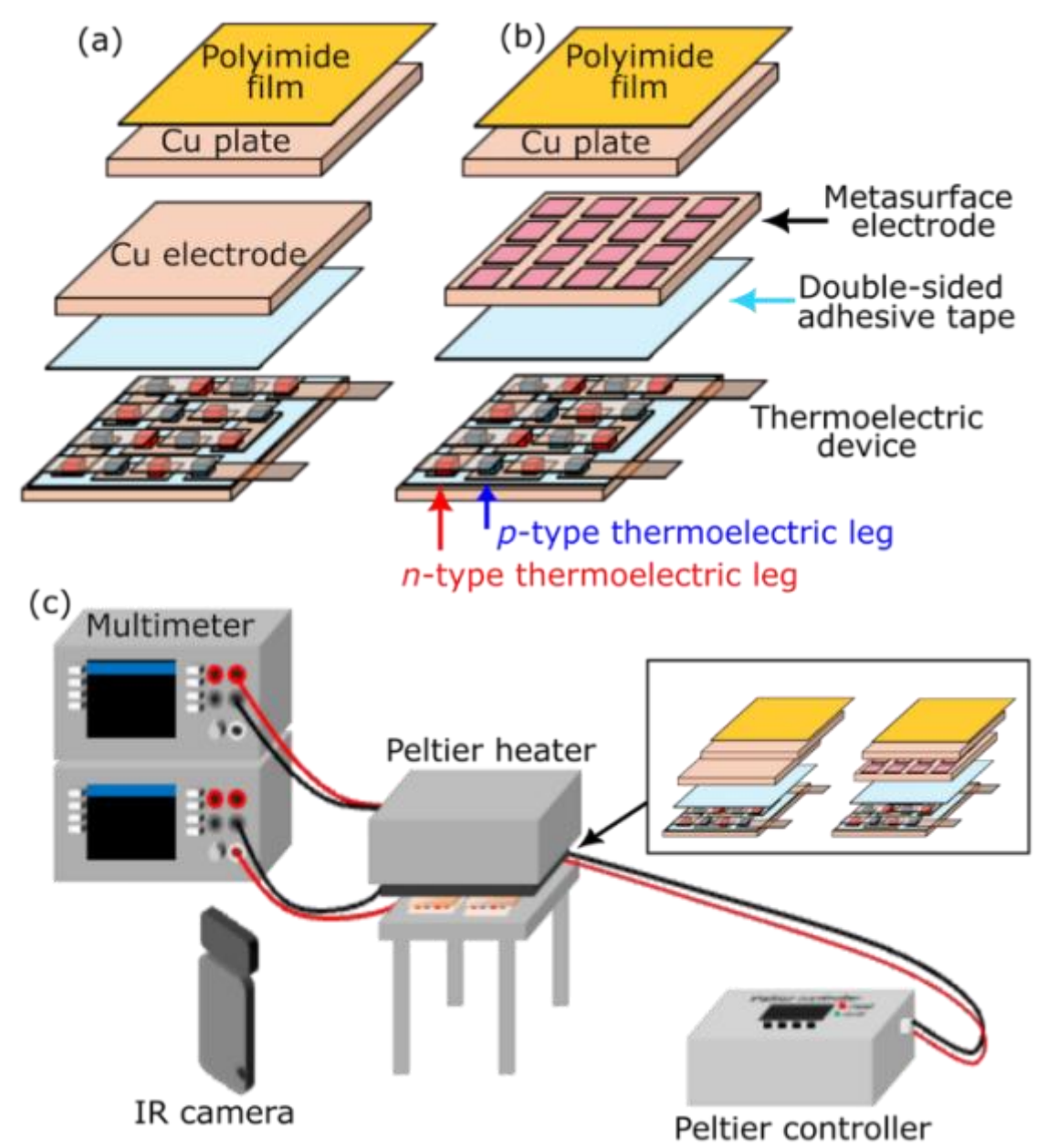

Figure 2. Exploded views of (a) the control thermoelectric device and (b) the metasurface device. (c) Experimental setup used for direct visualization of the heat extraction process.

Figure 3(a) shows the normalized output voltages of the metasurface and control thermoelectric devices as a function of time. When the heating started, the output voltages of both devices increased rapidly and reached nearly steady values of approximately 38 mV within 10-15 s, reflecting the establishment of a temperature difference across the thermoelectric modules.[25-28] The devices were then maintained under heating until approximately 55 s. After the heater was removed, the output voltages of both devices decreased as the Cu plates cooled, while a clear difference between the two voltage responses emerged during the initial stage of cooling.

To show the difference between the two devices, Fig. 3(b) plots the differential output voltage between the metasurface device and the control device ( $\Delta V = V_{MS} - V_c$ ). Immediately after the onset of cooling, the differential output voltage increased rapidly for the first 3 s, indicating enhanced thermoelectric output from the metasurface device. After 40 s, the differential output voltage became negative, indicating that the output voltage of the metasurface device fell below that of the control device. We infer that the MS-Electrode heats less efficiently and cools more efficiently than the planar Cu reference electrode.

Figures 3(c)(i-viii) show side-by-side infrared images of the Cu plates on the MS and control devices at 10-s intervals from 0 s to 70 s after the onset of cooling (see also Movie S1 in the Supplemental Material). Immediately after the onset of cooling, the temperatures of the Cu plates on the MS and the control devices were nearly identical at 384 K and 386 K, respectively. As cooling proceeded, a clear temperature difference developed. At 10 s, the temperatures were 355 K and 358 K, respectively, corresponding to a temperature difference of 3.0 K. These infrared images directly demonstrate metasurface-induced cooling enhancement, which was reproducibly observed over four independent measurements (Supplemental Material, Fig. S1).

The temperature difference peaked at 10 s and subsequently decreased with time (Fig. 3(b)). The Cu plate placed on the MS-device was actually 3.0 K cooler than that on the control device. The temporal evolution of the temperature difference explains the differential voltage behavior shown in Fig. 3(b). During the initial stage of cooling, the larger temperature drop of the Cu plate on the MS-device led to enhanced thermoelectric output. As the temperature difference across the thermoelectric element diminished, the output voltage of the metasurface device eventually became smaller than that of the control device.

To exclude the possibility that this asymmetric cooling originated from an intrinsic difference between the thermoelectric devices, we performed a control experiment using the same devices equipped with planar Cu electrodes. No appreciable difference in cooling was observed, confirming that the enhanced cooling originates from the metasurface electrode rather than intrinsic device-to-device variation (Movie S2 and Supplemental Material, Fig. S2).

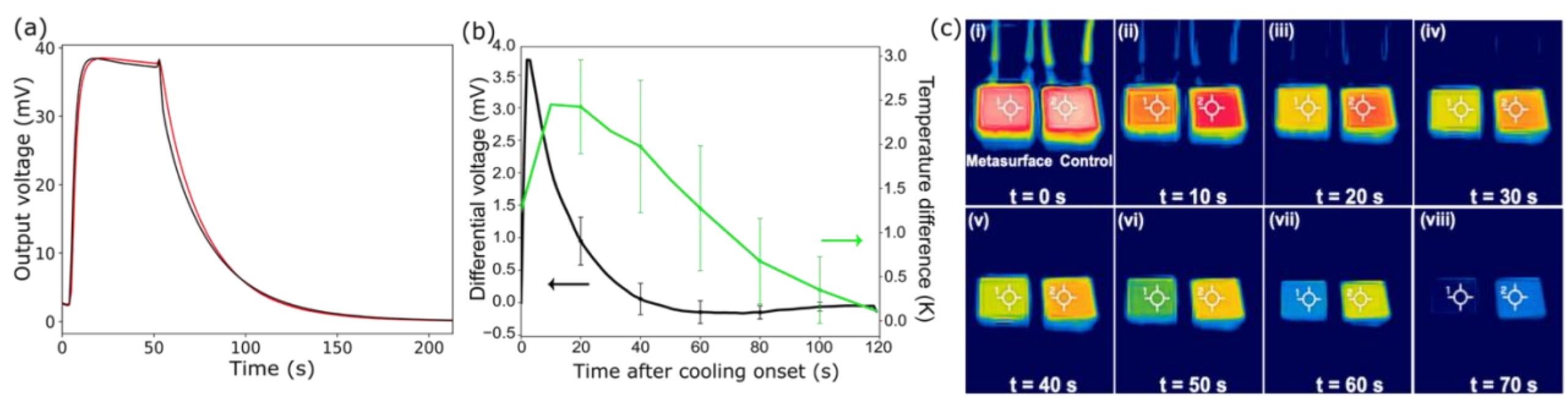

Figure 3. (a) Temporal evolution of the output voltages of the metasurface device (red line) and the control device (black line). (b) Differential output voltage (left axis) and the temperature difference (right axis) between the metasurface and control devices. When the Cu plate placed on the MS-device is cooler than that on the control device, the temperature difference is negative. The curves represent the mean values obtained from four independent measurements, with error bars indicating the standard deviation at 20-s intervals. (c) Side-by-side infrared images of the Cu plates placed on the metasurface (left) and the control (right) thermoelectric devices during the cooling process. The images were acquired at 10-s intervals from 0 s to 70 s after the onset of cooling.

Importantly, we infer from these results that relative to the planar Cu electrode, the MS-Electrode liberates additional thermal energy via radiative transfer. Under ordinary conditions, heat exchange between contacting solids is governed by thermal conduction.[29,30] However, the 100-nm-high Ag disks of the MS-Electrode produce a larger surface topography than the planar Cu electrode, and consequently the thermal boundary resistance (TBR) between the Cu plate and the metasurface electrode is larger than that of the planar electrode. This TBR limited the MS electrode cooling rate. Regardless of this limited cooling, we still find that the MS-Electrode exhibits a larger thermoelectric output, supporting an additional radiative heat-transfer channel.

Interestingly, the temperature difference is maximum shortly after the voltage difference is maximum and approximately when the voltage difference is decreasing most rapidly (Supplementary Fig. S3). Indeed, the differential voltage is more strongly related to the rate of change of the temperature difference, $\frac{d\Delta T}{dt}$, than to $\Delta T$ alone (Fig. S3). This temporal relationship likely arises from the time required for heat to diffuse between the Cu electrode and the thermoelectric element, resulting in a delay between the measured temperature of the Cu electrode and the local temperature gradient that generates the thermoelectric voltage.

To examine whether the observed cooling is consistent with electromagnetic absorption in the metasurface, we compared the heat extracted from the Cu plate ( $Q_{\mathrm{Cu}}$) with the heat absorbed by the metasurface ($Q_{\mathrm{MS}}$). The heat extracted from the Cu plate, $Q_{\mathrm{Cu}}$ , was estimated from the measured temperature difference according to

$$Q_{\mathrm{Cu}} = m_{\mathrm{Cu}} \cdot c_{\mathrm{Cu}} \cdot \Delta T, \qquad (1)$$

where $m_{\mathrm{Cu}}$, $c_{\mathrm{Cu}}$, and $\Delta T$ denote the mass, specific heat capacity, and the temperature change of the Cu plate, respectively. The resulting value of $Q_{\mathrm{Cu}}$ was $0.07 \pm 0.00$ J.

The heat absorbed by the metasurface was evaluated by finite-element analysis (COMSOL Multiphysics) using the volumetric Joule heating density,

$$q_{\mathrm{MS}} = \frac{1}{2}\mathrm{Im}(\varepsilon(\omega))\varepsilon_0|\mathbf{E}(\mathbf{r})|^2, \qquad (2)$$

where $\varepsilon_0$ is the vacuum permittivity, $\varepsilon(\omega)$ is the relative permittivity of the material, and $\mathbf{E}(\mathbf{r})$ is the electric field amplitude.

$$Q_{\mathrm{MS}} = q_{\mathrm{MS}} \cdot v_{\mathrm{MS}} \cdot t, \qquad (3)$$

where $v_{\mathrm{MS}}$ and $t$ represent the volume of the metasurface and the cooling time, respectively. The calculated heat power density was $1.4 \times 10^8$ W m$^{-3}$, yielding $Q_{\mathrm{MS}}$ = 0.08 J over 40 s. The good agreement between $Q_{\mathrm{Cu}}$ and $Q_{\mathrm{MS}}$ supports the interpretation that the enhanced cooling is associated with electromagnetic absorption by the metasurface.

Finally, we estimated the energy $Q_{\mathrm{power}}$ associated with thermoelectric conversion. The effective output power was integrated over the first 40 s after the onset of cooling to obtain the total electrical energy,

$$Q_{\mathrm{power}} = \int_0^{40} p_{\mathrm{TE}}\, dt, \qquad (4)$$

where $p_{\mathrm{TE}}$ denotes the output power of the thermoelectric device. The energy involved in thermoelectric conversion ($Q_{\mathrm{TE}}$) was then estimated from

$$Q_{\mathrm{TE}} = \frac{Q_{\mathrm{power}}}{\eta}, \qquad (5)$$

where $\eta$ denotes the thermoelectric conversion efficiency, given by

$$\eta = \frac{T_H - T_L}{T_H} \times \frac{\sqrt{1+zT}-1}{\sqrt{1+zT}+\frac{T_{\mathrm{L}}}{T_{\mathrm{H}}}}, \qquad (6)$$

where $T_{\mathrm{H}}$ and $T_{\mathrm{L}}$ are the temperatures of the hot and cold sides of the thermoelectric element, respectively. Here, $z$ denotes the thermoelectric figure of merit, and $T = \frac{T_{\mathrm{H}}+T_{\mathrm{L}}}{2}$ is the average temperature. Using $T_{\mathrm{H}} = 383$ K, $T_{\mathrm{L}} = 303$ K and $z = 6.2 \times 10^{-4}\ \mathrm{K}^{-1}$, the thermoelectric conversion efficiency was estimated to be $1.1 \times 10^{-2}$. Consequently, the energy involved in thermoelectric conversion was $Q_{\mathrm{TE}} = (0.07 \pm 0.00)$ J. Notably, $Q_{\mathrm{Cu}} = 0.07\,J$ , $Q_{\mathrm{MS}} = 0.08\,J$ , and $Q_{\mathrm{TE}} = 0.07\,J$ are all of comparable magnitude. Together with the infrared observations, this energy balance supports heat extraction from the Cu plate through electromagnetic absorption by the metasurface and subsequent thermoelectric transport.

In summary, we directly visualized the cooling of a Cu plate in contact with a metasurface-integrated thermoelectric device by infrared thermography. The Cu plate on the metasurface device exhibited enhanced cooling and larger thermoelectric output compared with the control device. Energy analysis showed comparable magnitudes of the heat extracted from the Cu plate, the energy absorbed by the metasurface, and the energy transported through thermoelectric conversion, supporting the role of electromagnetic absorption in the observed cooling.

More generally, the present study provides direct evidence of heat extraction associated with electromagnetic absorption, a thermal process that cannot be accessed by conventional far-field measurements. Our findings demonstrate that the effective heat extraction can be achieved even when the metasurface is in direct contact with the object being cooled, providing insight into nonequilibrium thermal processes and offering opportunities for thermal management based on metasurface nonradiative cooling.

## AUTHOR INFORMATION

**Corresponding Author**

*W.K. (w-kubo@cc.tuat.ac.jp)

**Author Contributions**

W. K. conceived the idea of this work. T. T. was responsible for the numerical simulations. W. K. and R. S. supervised the experiments, and S. W. conducted the experimental work. All authors contributed to data interpretation and manuscript preparation.

**Funding Sources**

W. Kubo thanks The Kurata Grants by The Hitachi Global Foundation (Grant number: 1574). This work was supported by JSPS KAKENHI Grant Number JP24H02232.

REFERENCES

[1] A. C. Overvig, S. A. Mann, and A. Alù, Phys. Rex. X **11**, 021050 (2021).

[2] J. Yao, R. Lin, M. K. Chen, and D. P. Tsai, Adv. Photonics **5**, 024001 (2023).

[3] A. Kristensen, J. K. W. Yang, S. I. Bozhevolnyi, S. Link, P. Nordlander, N. J. Halas, and N. A. Mortensen, Nat. Rev. Mater. **2**, 16088 (2016).

[4] A. Boltasseva and V. M. Shalaev, Metamaterials **2**, 1 (2008).

[5] A. Sakurai, K. Yada, T. Simomura, S. Ju, M. Kashiwagi, H. Okada, T. Nagao, K. Tsuda, and J. Shiomi, ACS Cent. Sci. **5**, 319 (2019).

[6] P. Bermel, M. Ghebrebrhan, M. Harradon, Y. X. Yeng, I. Celanovic, J. D. Joannopoulos, and M. Soljacic, Nanoscale Res. Lett. **6**, 549 (2011).

[7] S. Inampudi, J. Cheng, M. M. Salary, and H. Mosallaei, J. Opt. Soc. Am. B **35**, 39 (2018).

[8] C. Arnold, F. Marquier, M. Garin, F. Pardo, S. Collin, N. Bardou, J.-L. Pelouard, and J.-J. Greffet, Phys. Rev. B **86**, 035316 (2012).

[9] Y.-S. Duh *et al.*, Nat. Commun. **11**, 4101 (2020).

[10] Q. Chu, F. Zhong, X. Shang, Y. Zhang, S. Zhu, and H. Liu, Nanophotonics **13**, 1279 (2024).

[11] G. Baffou, F. Cichos, and R. Quidant, Nat. Mater. **19**, 946 (2020).

[12] G. Baffou, P. Bon, J. Savatier, J. Polleux, M. Zhu, M. Merlin, H. Rigneault, and S. Monneret, ACS Nano **6**, 2452 (2012).

[13] Z. Chen, X. Shan, Y. Guan, S. Wang, J.-J. Zhu, and N. Tao, ACS Nano **9**, 11574 (2015).

[14] G. Baffou, C. Girard, and R. Quidant, Phys. Rev. Lett. **104**, 136805 (2010).

[15] N. Kawamura, T. Tanaka, and W. Kubo, ACS Photonics **11**, 1221 (2024).

[16] S. Katsumata, T. Tanaka, and W. Kubo, Opt. Express **29**, 16396 (2021).

[17] R. Nakayama, S. Saito, T. Tanaka, and W. Kubo, Nanophotonics **13**, 1361 (2024).

[18] S. Saito, A. Yamamoto, Y.-J. Lu, T. Tanaka, and W. Kubo, Discov. Nano **20**, 44 (2025).

[19] S. Hirobe, S. Wredh, J. K. W. Yang, and W. Kubo, Appl. Therm. Eng. **256**, 124080 (2024).

[20] K. Hamada, H.-H. Hsiao, and W. Kubo, Sci. Rep. **14**, 31842 (2024).

[21] M. L. Brongersma, N. J. Halas, and P. Nordlander, Nat. Nanotechnol. **10**, 25 (2015).

[22] W. Kubo, M. Kondo, and K. Miwa, J. Phys. Chem. C **123**, 21670 (2019).

[23] N. I. Landy, S. Sajuyigbe, J. J. Mock, D. R. Smith, and W. J. Padilla, Phys. Rev. Lett. **100**, 207402 (2008).

[24] T. Cao, C.-w. Wei, R. E. Simpson, L. Zhang, and M. J. Cryan, Sci. Rep. **4**, 3955 (2014).

[25] R. He, G. Schierning, and K. Nielsch, Adv. Mater. Technol. **3**, 1700256 (2018).

[26] Fitriani, R. Ovik, B. D. Long, M. C. Barma, M. Riaz, M. F. M. Sabri, S. M. Said, and R. Saidur, Renew. Sust. Energ. Rev. **64**, 635 (2016).

[27] F. Tohidi, S. Ghazanfari Holagh, and A. Chitsaz, Appl. Therm. Eng. **201**, 117793 (2022).

[28] X. F. Zheng, C. X. Liu, Y. Y. Yan, and Q. Wang, Renew. Sust. Energ. Rev. **32**, 486 (2014).

[29] T. N. Narasimhan, Rev. Geophys. **37**, 151 (1999).

[30] X. Tian, X. Lai, P. Zhu, and L. Wang, Proc. R. Soc. A: Math. Phys. Eng. Sci. **472**, 20160362 (2016).

Supplementary information

# Direct Visualization of Non-equilibrium Heat Extraction by a Metasurface Absorber

Shu Watanabe,[1] Robert E. Simpson,[2] Takuo Tanaka,[3-6] and Wakana Kubo[1]*†

[1] *Department of Electrical Engineering and Computer Science, Tokyo University of Agriculture and Technology, 2-24-16 Naka-cho, Koganei-shi, Tokyo 184-8588, Japan*
[2] *School of Engineering, University of Birmingham, Edgbaston, Birmingham, B15 2TT, UK.*
[3] *Innovative Photon Manipulation Research Team, RIKEN Center for Advanced Photonics, Wako, Saitama 351-0198, Japan.*
[4] *Metamaterials Laboratory, RIKEN Cluster for Pioneering Research, Wako, Saitama 351-0198, Japan.*
[5] *Department of Physics, Faculty of Science, Gakushuin University, Tokyo 171-8588, Japan.*
[6] *Institute of Post-LED Photonics, Tokushima University, Tokushima 770-8560, Japan.*

Supplementary material 1 (SM1): Movie temperature distribution of the Cu plates placed on the MS and control devices.

Supplementary material 2 (SM2) Movie temperature distribution of the Cu plates placed on the control devices.

Supplementary figures: S1-S3.

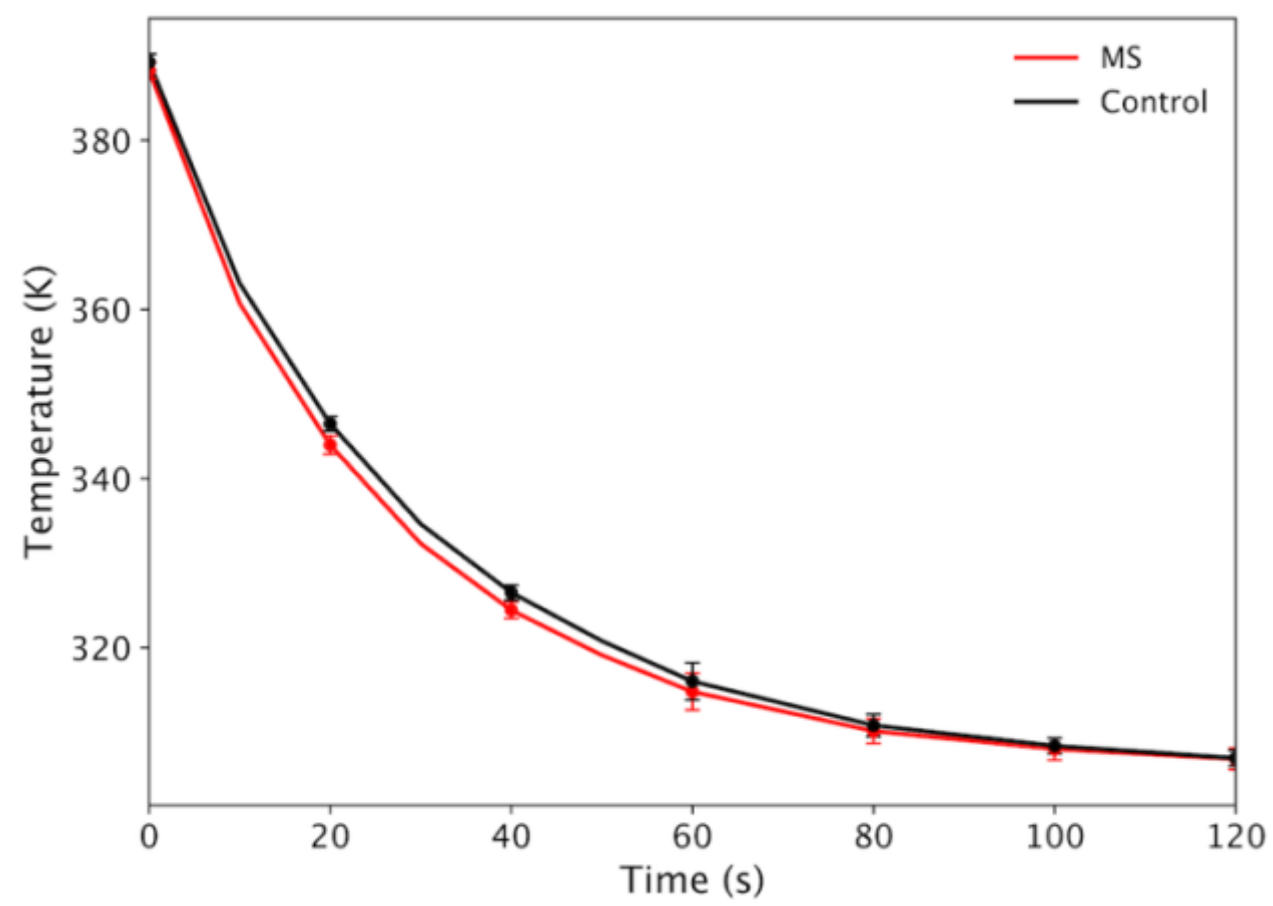

Figure S1. Temperature evolution of the Cu plates placed on the metasurface (MS: red) and control (black) thermoelectric devices during cooling. The red and black curves show the mean temperatures of the Cu plates placed on the MS and control devices, respectively, obtained from four independent measurements. Error bars, shown at 20-s intervals for clarity, represent the standard deviation of four independent measurements.

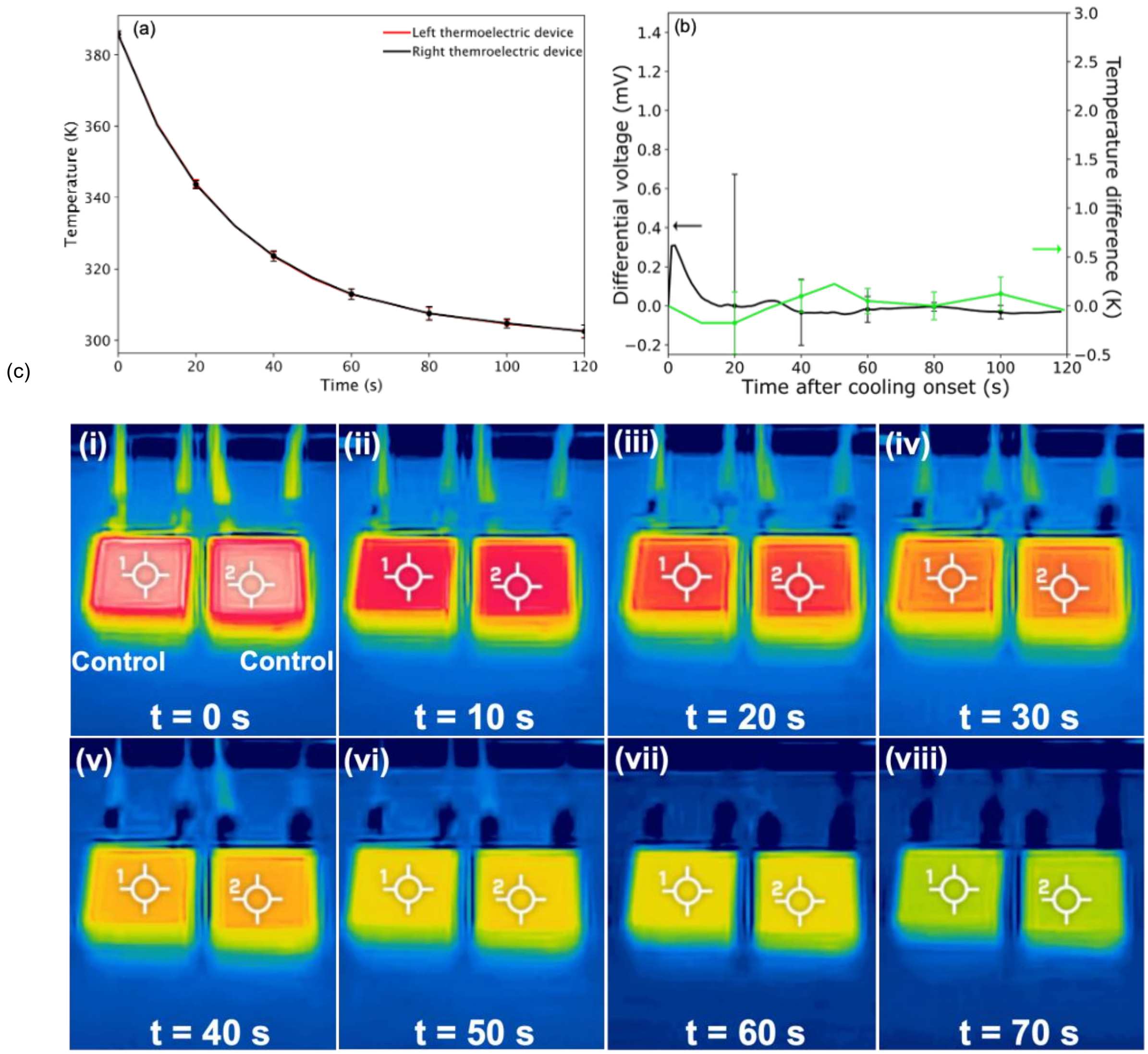

Figure S2. Control experiment using two control thermoelectric devices, with a Cu plate placed on each device. (a) Temperature evolution of the Cu plates placed on the left (red) and right (black) control devices during cooling. The curves represent the mean temperatures obtained from four independent measurements. Error bars, shown at 20-s intervals for clarity, represent the standard deviation of four independent measurements. (b) Differential output voltage (black line, left axis) and temperature difference (green line, right axis) between the two control devices during cooling. The curves represent the mean values obtained from four independent measurements, with error bars indicating the standard deviation at 20-s intervals. (c) Side-by-side infrared images of the Cu plates placed on the two control thermoelectric devices. The images were acquired at 10-s intervals from 0 to 70 s after the onset of cooling. The difference in the overall color appearance between these infrared images and those in Fig. 3(c) results from the automatic adjustment of the temperature color scale by the infrared camera; therefore, the colors should not be directly compared between the two measurements.

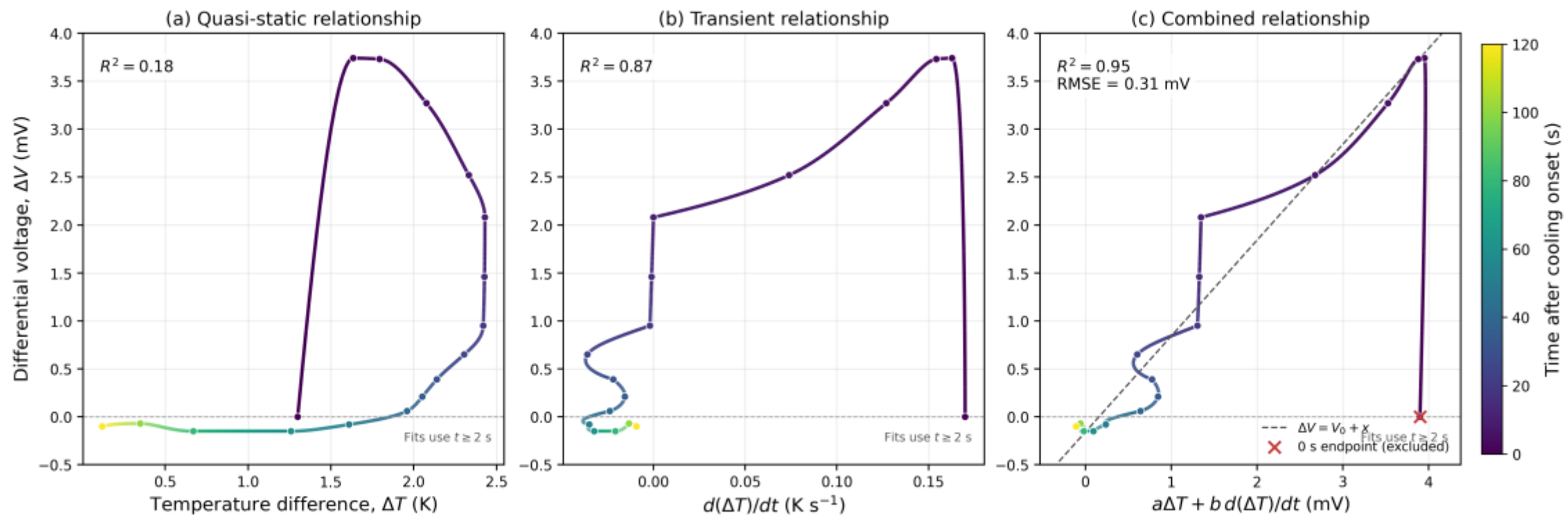

Figure S3. Relationship between the differential voltage and temperature response during cooling. The differential voltage $\Delta V$ is compared with (a) the temperature difference $\Delta T$ (quasi-static relationship), (b) its time derivative $d(\Delta T)/dt$ (transient relationship), and (c) their combined contribution $a\Delta T + b\, d(\Delta T)/dt$. The corresponding coefficients of determination are $R^2 = 0.18$, 0.87, and 0.95, respectively. The combined model gives $R^2 = 0.95$ and an RMSE of 0.31 mV. The color scale indicates the time after the onset of cooling, and the data at $t = 0$ s were excluded from the fitting.